\pdfoutput=1
\documentclass{iopjournal}

\usepackage{cite}
\usepackage{amsmath,amssymb}
\usepackage{bm}
\usepackage{microtype}
\hypersetup{hypertexnames=false}

\newcommand{\vac}{\,{\raisebox{0.15ex}{\setlength{\fboxsep}{0pt}\setlength{\fboxrule}{0.45pt}\fbox{\phantom{\rule{0.82ex}{0.82ex}}}}}\,}

\begin{document}

\articletype{Paper}

\title{Macroscopic nuclear resonance energy transfer mediated by coherent magnons}

\author{Ryan Weed$^{1,*}$} 

\affil{$^1$Eos Atomics, San Francisco, CA 94107, USA}

\affil{$^*$Author to whom any correspondence should be addressed.}

\email{ryan.weed@eosatomics.com}


\begin{abstract}
Phonon-mediated nuclear resonance energy transfer (RET) in metal deuterides is severely constrained by nuclear rest-mass dilution, mechanical lattice fracture limits, and thermal nuclear spin cancellation. Here we extend the collective nuclear Dicke model to coherent electron spin waves (magnons) driven at ferromagnetic resonance (FMR) in a cryogenically cooled ferromagnetic palladium--nickel deuteride ($\mathrm{Pd}_{1-x}\mathrm{Ni}_x\mathrm{D}_y$) lattice. Replacing mechanical ion displacement with local Fermi-contact and core-polarization hyperfine fields eliminates nuclear rest-mass dilution via Turov--Petrov hyperfine-amplified magnetic dipole (M1) coupling between defect-screened $\mathrm{D}_2$ donor pairs and prominent spin-flip M1 giant resonances in $^{106}\mathrm{Pd}$ and $^{58}\mathrm{Ni}$ host nuclei. Incorporating Feshbach channel-closing doorway asymmetry and Suhl four-magnon spin-wave saturation into a finite-time Lippmann--Schwinger off-shell dressing integral yields a pre-asymptotic volumetric scaling law for the energy-transfer rate, $\Gamma_{\mathrm{transfer}} \propto f_M^{-10.8}$, across FMR drive frequency $f_M$ above the low-field domain-saturation rollover. Whereas multi-acceptor transfers populate on-shell states below host nucleon separation thresholds -- yielding either radiationless lattice thermalization or a falsifiable FMR-gated $3\text{--}9~\mathrm{MeV}$ M1 gamma-ray cascade -- rare single-acceptor excitations predict monoenergetic two-body breakup endpoints at $14.15~\mathrm{MeV}$ ($^{106}\mathrm{Pd}$) and $11.43~\mathrm{MeV}$ ($^{58}\mathrm{Ni}$) for neutrons and $19.87~\mathrm{MeV}$ for alpha particles, alongside strongly suppressed $2.45~\mathrm{MeV}$ neutron leakage, reconciling recent sub-keV ion-beam and underground thermal fusion benchmarks.
\end{abstract}

\section{Introduction}
\label{sec:intro}

The extreme discrepancy between the spontaneous $\mathrm{D}_2$ fusion rate in vacuum ($\Gamma \approx 3 \times 10^{-64}~\mathrm{s}^{-1}$ per pair) \cite{Koonin1989} and enhanced reaction yields reported in highly loaded metal deuterides ($\Gamma \approx 10^{-23}\text{--}10^{-12}~\mathrm{s}^{-1}$) \cite{Kasagi2002,Raiola2002,Huke2008,Czerski2016,Metzler2024} cannot be accounted for by bare two-body Gamow tunneling. Recently, Karahadian \textit{et al} \cite{Karahadian2026} and Chen \textit{et al} \cite{Chen2025} combined electrochemical loading with sub-keV deuteron bombardment ($E_{\mathrm{cm}} = 0.25\text{--}6.5~\mathrm{keV}$) in $\mathrm{Pd}$ and $\mathrm{Ti}$ foils, revealing a flat reaction-yield plateau below $2~\mathrm{keV}$ exceeding bare-nucleus expectations by $10^{18}$, with effective screening potentials reaching $U_e = 0.8\text{--}3.0~\mathrm{keV}$ in defect-rich surface layers \cite{Karahadian2026,Cvetinovic2023,Dubey2025}. Furthermore, underground measurements by Lipoglav\v{s}ek and Cvetinovi\'{c} \cite{Lipoglavsek2026} demonstrated spontaneous $\mathrm{D\text{--}D}$ fusion at purely thermal energy ($E = 0.025~\mathrm{eV}$, $T = 289~\mathrm{K}$) in cold-rolled $\mathrm{Pd}$ foils without ion bombardment, detecting $3.0~\mathrm{MeV}$ protons originating from subsurface $^{19}\mathrm{F}$ and $^{16}\mathrm{O}$ defect traps ($2.0\text{--}7.0~\mu\mathrm{m}$ depth, $U_e = 475 \pm 10~\mathrm{eV}$). Crucially, their windowless $\Delta E\text{--}E$ coincidence measurements in a cosmic-ray-suppressed mercury mine ruled out the $5:1$ internal $e^-e^+$ pair-emission signature predicted for a narrow $J^\pi = 0^+$ threshold resonance in $^4\mathrm{He}^*$ \cite{Czerski2016,Dubey2025,Lipoglavsek2026}, proving that thermal solid-state fusion cannot proceed via isolated internal lepton-pair de-excitation.

\begin{figure}[htbp]
\centering
\includegraphics[width=\textwidth]{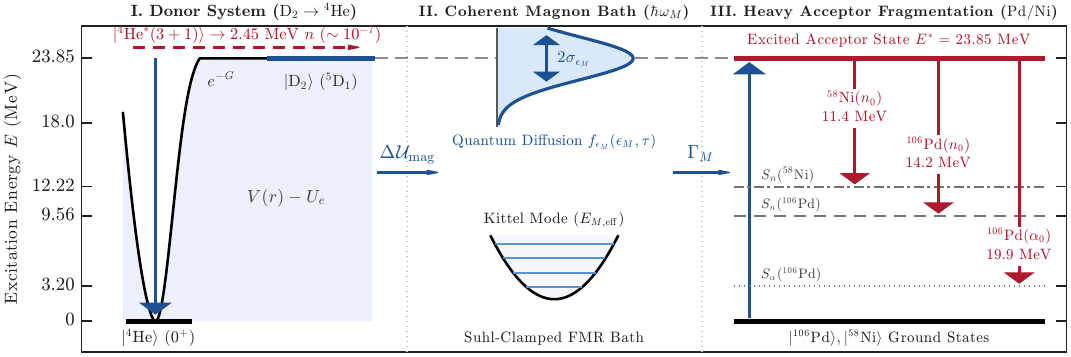}
\caption{Calibrated three-zone nuclear energy-level and quantum pathway diagram for FMR-driven Magnon Resonance Energy Transfer (Magnon-RET) across $E = 0\text{--}26~\mathrm{MeV}$. (I)~\textbf{Donor ($\mathrm{D}_2 \to {}^4\mathrm{He}$):} $|\mathrm{D}_2\rangle$ ($^5\mathrm{D}_1$) pairs at $\Delta M c^2 = 23.85~\mathrm{MeV}$ (see appendix~\ref{app:dirac} for the secondary $^3\mathrm{P}_1$ channel) tunnel through the defect-screened barrier $V(r) - U_e$ ($e^{-G}$) to $|{}^4\mathrm{He}\rangle$ ($0^+$). Rapid Gamow decay ($\hbar\gamma_{\mathrm{tunnel}} = 5.95~\mathrm{MeV}$) of the open $|{}^4\mathrm{He}^*(3+1)\rangle$ Feshbach doorway resonance ($24.25~\mathrm{MeV}$) breaks destructive pathway interference ($\vert{}1-\eta\vert{} \approx 0.1$) while emitting trace $2.45~\mathrm{MeV}$ neutrons ($\Gamma_{3+1}/\Gamma_{\mathrm{transfer}} \sim 10^{-7}$ at $915~\mathrm{MHz}$). (II)~\textbf{Magnon Bath ($\hbar\omega_M$):} Collective coupling $\Delta\mathcal{U}_{\mathrm{mag}}$ drives the Suhl-clamped $k=0$ Kittel condensate ($E_{M,\mathrm{eff}}$), whose rapid collective exchange ($\Gamma_M^{\mathrm{eff}}$) dresses intermediate doorway states via Gaussian quantum diffusion $f_{\epsilon_M}(\epsilon_M,\tau)$ of full width $2\sigma_{\epsilon_M}$ before projecting onto the on-shell $23.85~\mathrm{MeV}$ final manifold. (III)~\textbf{Heavy Acceptor Fragmentation ($\mathrm{Pd/Ni}$):} On-shell single-acceptor ($m=1$) excitation of $|{}^{106}\mathrm{Pd}\rangle$ and $|{}^{58}\mathrm{Ni}\rangle$ to $E^* = 23.85~\mathrm{MeV}$ above $S_n(^{58}\mathrm{Ni}) = 12.22~\mathrm{MeV}$ (dash-dotted), $S_n(^{106}\mathrm{Pd}) = 9.56~\mathrm{MeV}$ (dashed), and $S_\alpha(^{106}\mathrm{Pd}) = 3.20~\mathrm{MeV}$ (dotted) drives two-body breakup ($Q_0 = E^* - S_{n,\alpha}$), yielding $11.4~\mathrm{MeV}$ ($^{58}\mathrm{Ni}$ $n_0$) and $14.2~\mathrm{MeV}$ ($^{106}\mathrm{Pd}$ $n_0$) neutrons and $19.9~\mathrm{MeV}$ ($^{106}\mathrm{Pd}$ $\alpha_0$) $\alpha$ particles.}
\label{fig:energy_pathway}
\end{figure}

Collective solid-state nuclear models based on Dicke superradiance \cite{Dicke1954} propose transferring the $\Delta M c^2 = 23.85~\mathrm{MeV}$ mass defect of $|\mathrm{D}_2\rangle \to |{}^4\mathrm{He}\rangle$ non-radiatively to heavy host nuclei ($^{102\text{--}110}\mathrm{Pd}$ or $^{58\text{--}64}\mathrm{Ni}$) via a shared $k=0$ phonon bath \cite{Metzler2024,Hagelstein2023}. However, phonon-mediated resonance energy transfer (RET) faces severe physical bottlenecks. First, lowest-order relativistic phonon-nuclear boost coupling ($\hat{H}_{\mathrm{int}}^{\mathrm{phonon}} = \mathbf{a}\cdot c\hat{\mathbf{P}}_j$) \cite{Hagelstein2023} requires accelerating the nuclear rest mass $M_j$ of all $N$ ions in coherent volume $V_{\mathrm{eff}} = N v_{\mathrm{cell}}$. The single-ion boost velocity $\beta_j = \sqrt{2E_A / (N M_j c^2)} = \sqrt{2u_A / (\rho_m c^2)}$ is diluted by the crystal rest-energy density ($M_{\mathrm{Pd}}c^2 \approx 99.1~\mathrm{GeV\,atom^{-1}}$, $\rho_m c^2 \approx 1.1 \times 10^{21}~\mathrm{J\,m^{-3}}$) and capped by mechanical fracture strain ($\epsilon_{\mathrm{max}} \sim 10^{-3}$, clamping $u_{A,\mathrm{max}} \approx 5.7 \times 10^4~\mathrm{J\,m^{-3}}$ and $\beta_j \lesssim 10^{-8}$). Second, the odd-parity boost operator is proportional to the nuclear spin projection $M_J \in \{-1, 0, +1\}$, which averages to zero ($\langle M_J \rangle = 0$) in an unpolarized thermal lattice while exciting hindered odd-parity multipoles rather than prominent low-lying magnetic dipole strength (appendix~\ref{app:dirac}).

In this paper, we extend the nuclear Dicke model to coherent electron spin waves (magnons) driven at ferromagnetic resonance (FMR) in a cryogenically cooled $\mathrm{Pd}_{1-x}\mathrm{Ni}_x\mathrm{D}_y$ alloy (figure~\ref{fig:energy_pathway}). We show that FMR pumping eliminates inertial mass dilution via Turov--Petrov hyperfine-amplified magnetic dipole (M1) coupling \cite{TurovPetrov1972,deGennes1963}, directly addresses collective $0^+ \to 1^+$ spin-flip giant resonances in $\mathrm{Pd}$ and $\mathrm{Ni}$ \cite{Heyde2010,Kopecky1987,Brandherm2024}, establishes a volumetric frequency scaling law bounded by Suhl's non-linear spin-wave instability \cite{Suhl1957}, and predicts falsifiable two-body breakup and radiative signatures that reconcile recent sub-keV and underground solid-state fusion benchmarks.

\section{Turov--Petrov magnon-nuclear Hamiltonian}
\label{sec:hamiltonian}

In the many-body Dirac Foldy--Wouthuysen representation, the lowest-order relativistic phonon-nuclear interaction and center-of-mass ion momentum operator are \cite{Hagelstein2023}
\begin{subequations}
\label{eq:phonon_all}
\begin{align}
\hat{H}_{\mathrm{int}}^{\mathrm{phonon}} &= \mathbf{a}\cdot c\hat{\mathbf{P}}_j = -i\frac{E_2 - E_1}{2 M_j c^2}\left( \sum_{n=1}^{A}\bm{\sigma}_n \times \frac{\hat{\bm{\pi}}_n}{m_n c} \right) \cdot c\hat{\mathbf{P}}_j, \label{eq:phonon_int} \\
\hat{\mathbf{P}}_j &= \mathbf{e}_j^{(A)}\sqrt{\frac{\hbar M_j \omega_A}{2N}}\left( \frac{\hat{a}_A - \hat{a}_A^\dagger}{i} \right). \label{eq:phonon_p}
\end{align}
\end{subequations}
Because $\langle \hat{\mathbf{P}}_j^2 \rangle_{\mathrm{osc}} = M_j E_A / N$ distributes a fixed acoustic energy $E_A$ across the total lattice mass, the boost factor $\sqrt{E_A / (N M_j c^2)}$ incurs an explicit $1/\sqrt{N}$ dilution that cancels the macroscopic Dicke factor $\sqrt{N_{\mathrm{D}_2}N_{\mathrm{Rec}}}$ at fixed input power.

Conversely, magnons are phase-locked precessions of polarized $d$- and $s$-band electron spins that transfer angular momentum to embedded nuclei via local Fermi-contact and core-polarization hyperfine coupling $\hat{H}_{\mathrm{hf}} = \sum_k A_{\mathrm{hf},k}\hat{\mathbf{I}}_k \cdot \hat{\mathbf{S}}_e(\mathbf{r}_k)$ without displacing the nuclear center of mass \cite{TurovPetrov1972,deGennes1963}. Driving the uniform $k=0$ Kittel mode ($\hat{m}^\dagger, \hat{m}$) at in-plane FMR frequency $\omega_M = \gamma_e[B_{\mathrm{DC}}(B_{\mathrm{DC}} + \mu_0 M_s)]^{1/2}$ \cite{Kittel1948} with transverse radio-frequency (RF) field $h_{\mathrm{RF}} \equiv \mu_0 H_{\mathrm{RF}} \perp \mathbf{B}_{\mathrm{DC}}$ tilts the site-specific static hyperfine field $B_{\mathrm{hf},k}$ ($10\text{--}35~\mathrm{T}$ at $\mathrm{Pd}/\mathrm{Ni}$ host sites via core polarization; $0.2\text{--}1.5~\mathrm{T}$ at interstitial $\mathrm{D}$ sites via $s\text{--}d$ hybridization) by the resonant precession angle $\theta_\perp = M_\perp / M_s = h_{\mathrm{RF}} / \Delta H$, where $\Delta H \equiv \mu_0\Delta H_{\mathrm{FMR}}$ is the FMR full-width at half-maximum linewidth \cite{TurovPetrov1972,deGennes1963,Suhl1958,Nakamura1958}. In the Turov--Petrov effective-field representation, the nuclear M1 Hamiltonian is
\begin{subequations}
\label{eq:magnon_ham}
\begin{align}
\hat{H} &= \hat{H}_{\mathrm{nuclei}} + \hbar\omega_M \hat{m}^\dagger \hat{m} - \sum_{j,j',k}\left( |\Phi_j\rangle\,\langle\Phi_j | \bm{\mu}_N \cdot \hat{\mathbf{B}}_{M,k} | \Phi_{j'}\rangle\,\langle\Phi_{j'}| \right)_k, \label{eq:ham_tot} \\
\hat{\mathbf{B}}_{M,k} &= \hat{\mathbf{e}}_\perp \frac{|B_{\mathrm{hf},k}|}{\sqrt{2 N_s S}}\left( \hat{m}^\dagger + \hat{m} \right), \label{eq:bm_op}
\end{align}
\end{subequations}
where $N_s$ is the number of polarized electron spins of spin $S$ in $V_{\mathrm{eff}}$. At host acceptor sites ($\vert{}B_{\mathrm{hf,Rec}}\vert{} \equiv \vert{}B_{\mathrm{hf},0}\vert{}$), the Turov--Petrov hyperfine enhancement factor $\mu_{r,\mathrm{eff}} \equiv B_{M,\mathrm{eff}} / h_{\mathrm{RF}} = |B_{\mathrm{hf},0}| / \Delta H \sim 10^3$ (for alloy-averaged $|B_{\mathrm{hf},0}| \approx 10~\mathrm{T}$ and $\Delta H = 10~\mathrm{mT}$) yields a mean-square dynamic nuclear hyperfine field
\begin{equation}
\langle \hat{\mathbf{B}}_M^2 \rangle_{\mathrm{osc}} = B_{M,\mathrm{eff}}^2 = \mu_{r,\mathrm{eff}}^2 h_{\mathrm{RF}}^2 \equiv \frac{2\mu_0 E_M}{V_{\mathrm{eff}}}. \label{eq:b_sq}
\end{equation}
Here $E_M$ (with intensive coupling density $u_M \equiv E_M / V_{\mathrm{eff}} = B_{M,\mathrm{eff}}^2 / (2\mu_0)$) parameterizes the hyperfine-amplified nuclear coupling scale, which exceeds the physical electromagnetic spin-precession energy stored in the sample by $\sim \mu_{r,\mathrm{eff}}^2$ because the static exchange-split $d$ band supplies $B_{\mathrm{hf},0}$ internally \cite{TurovPetrov1972}. Crucially, neither $M_j c^2$ nor mechanical fracture strain enters equation~(\ref{eq:b_sq}).

Because the nuclear magnetic moment $\bm{\mu}_N$ is a rank-1 positive-parity axial vector ($\Delta J = 1, \Delta\pi = \text{no}$), Wigner--Eckart triangle selection rules require a $J^\pi = 1^+$ donor configuration to reach the $|{}^4\mathrm{He}~0^+\rangle$ ground state. In free space, ground-state ortho-$\mathrm{D}_2$ ($S \in \{0, 2\}$) occupies $L = 0$, whereas the $L = 2$ rotational level lies $6B_{\mathrm{rot}} \approx 22.3~\mathrm{meV}$ higher. Inside an anisotropic superabundant-vacancy ($\mathrm{Pd}_3\vac\mathrm{D}_4$) defect cage \cite{Fukai1994}, however, non-cubic Jahn--Teller lattice relaxation breaks rotational isotropy ($[\hat{H}_{\mathrm{crys}}, \hat{\mathbf{L}}^2] \neq 0$), admixing $L = 2$ ($^5\mathrm{D}_1$, $J^\pi = 1^+, T = 0$) orbital character directly into the quintet ($S = 2$) ground state without thermal excitation (and permitting exchange-mixed $^3\mathrm{S}_1$ capture if the two deuterons occupy inequivalent sub-sites). This $1^+$ donor component couples directly to the roughly $15\%$ $^5\mathrm{D}_0$ admixture of $|{}^4\mathrm{He}~0^+\rangle$, where two-body meson-exchange currents lift single-nucleon isoscalar M1 suppression in low-energy $d(d,\gamma){}^4\mathrm{He}$ capture \cite{Langenbrunner1990,Tilley1992}:
\begin{equation}
U_{\mathrm{mag}}^2 = |\langle \mathrm{D}_2~{}^5\mathrm{D}_1 | \bm{\mu}_N \cdot \hat{\mathbf{B}}_{M,\mathrm{D}} | {}^4\mathrm{He}~0^+ \rangle|^2 = \left( \mathcal{M}_{\mathrm{M1}}\sqrt{\frac{v_{\mathrm{nuc}}}{v_{\mathrm{mol}}}}e^{-G_{5\mathrm{D}}} \right)^2 \mu_N^2 \left( \frac{2\mu_0 E_M}{V_{\mathrm{eff}}} \right), \label{eq:umag_sq}
\end{equation}
with nuclear-to-molecular volume ratio $v_{\mathrm{nuc}}/v_{\mathrm{mol}} = 6.64 \times 10^{-12}$ and nuclear magneton $\mu_N = 3.152 \times 10^{-8}~\mathrm{eV\,T^{-1}}$. Importantly, the WKB exponent $G_{5\mathrm{D}} = G_{L=0}(U_e) + \Delta G_{L=2} \approx 33.9$ explicitly includes the short-range $L = 2$ centrifugal barrier increment ($\Delta G_{L=2} \approx 7\text{--}10$ integrated from classical turning point $r_{\mathrm{tp}} = e^2/(4\pi\epsilon_0 U_e) \approx 1.0\text{--}1.25~\mathrm{pm}$ down to $R_{\mathrm{nuc}} \approx 4.3~\mathrm{fm}$) atop the bare $s$-wave Gamow exponent ($G_{L=0} \approx 23.4\text{--}27.0$ across $U_e \approx 1.15\text{--}1.5~\mathrm{keV}$). Similarly, the effective donor factor $\mathcal{M}_{\mathrm{M1}} \equiv \xi_{\mathrm{D}}\xi_{L=2}\tilde{\mathcal{M}}_{\mathrm{M1}}$ absorbs both the crystal-field $d$-wave amplitude $\xi_{L=2}$ and donor-to-host hyperfine ratio $\xi_{\mathrm{D}} = |B_{\mathrm{hf,D}}/B_{\mathrm{hf},0}| \sim 0.05\text{--}0.15$ (taking reference $\mathcal{M}_{\mathrm{M1}} \sim 10$ below; because $U_{\mathrm{mag}}$ enters equation~(\ref{eq:conv_closed}) only inside the double-logarithmic factor $\kappa \approx 6.8$ while host $|B_{\mathrm{hf},0}|$ governs quantum diffusion, varying $\mathcal{M}_{\mathrm{M1}}e^{-G_{5\mathrm{D}}}$ by two orders of magnitude shifts $\Gamma_{\mathrm{transfer}}$ by $<8\%$). On the heavy-acceptor side, M1 coupling directly excites collective $J^\pi = 0^+ \to 1^+$ spin-flip giant resonances ($1f_{7/2}^{-1}1f_{5/2}$ in $^{58}\mathrm{Ni}$ at $8.4\text{--}9.2~\mathrm{MeV}$ and $1g_{9/2}^{-1}1g_{7/2}$ in $^{106}\mathrm{Pd}$ centered near $8.8~\mathrm{MeV}$) \cite{Heyde2010,Kopecky1987,Brandherm2024}.

\section{Doorway asymmetry, Suhl clamping, and frequency scaling}
\label{sec:scaling}

Because single magnons ($\hbar\omega_M \sim \mu\mathrm{eV}$) cannot bridge $\Delta M c^2 = 23.85~\mathrm{MeV}$ at first order, energy transfer proceeds via L\"{o}wdin--Feshbach resolvent pathways between the $N_{\mathrm{D}_2}$ donor ensemble and $N_{\mathrm{Rec}}$ heavy-acceptor manifold \cite{Lowdin1951,Lowdin1962,Feshbach1958,Feshbach1962,Feshbach1967}. As proved in appendix~\ref{app:lowdin}, upper and lower virtual pathways cancel identically in a closed Hermitian subspace ($\eta = 1$). However, in the donor-first pathway the initial $A=4$ subsystem sits at $23.85~\mathrm{MeV}$ ($4.04~\mathrm{MeV}$ above the $19.81~\mathrm{MeV}$ $p+t$ breakup threshold), mixing ($W \approx 1.15~\mathrm{MeV}$) with the open $|{}^4\mathrm{He}^*(3+1)\rangle$ Feshbach doorway resonance at $24.25~\mathrm{MeV}$ ($\hbar\gamma_{\mathrm{tunnel}} \approx 5.95~\mathrm{MeV}$) \cite{Tilley1992,Feshbach1967}, whereas in the acceptor-first virtual state ($\Delta E_{\mathrm{Rec}} \approx \Delta M c^2$) the available $A=4$ subsystem energy $E_i - \Delta E_{\mathrm{Rec}} \approx 0$ lies $19.81~\mathrm{MeV}$ below threshold where $3+1$ continuum decay is kinematically closed. This threshold channel-closing asymmetry yields $|1-\eta| \approx 0.1$ and the Dicke-enhanced coupling
\begin{equation}
|\Delta\mathcal{U}_{\mathrm{mag}}|^2 = |1 - \eta|^2 U_{\mathrm{mag}}^2 N_{\mathrm{D}_2} N_{\mathrm{Rec}}. \label{eq:dicke_u}
\end{equation}

\begin{figure}[htbp]
\centering
\includegraphics[width=0.55\textwidth]{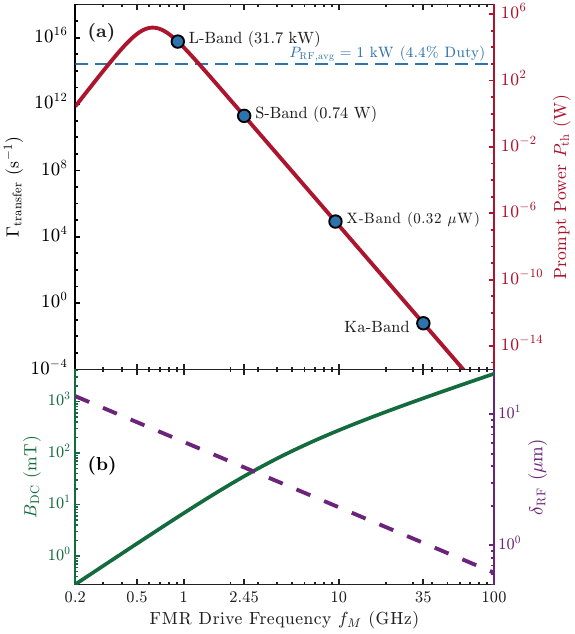}
\caption{Volumetric frequency scaling and Kittel electrodynamics for a Suhl-clamped $\mathrm{Pd}_{0.8}\mathrm{Ni}_{0.2}\mathrm{D}_x$ ($\mathrm{Pd}_3\vac\mathrm{D}_4$) target ($\mu_0 M_s = 0.18~\mathrm{T}$, $\Delta H = 10~\mathrm{mT}$, $h_{\mathrm{crit}} = 2.3~\mathrm{mT}$, $\mu_{r,\mathrm{eff}} = 10^3$, $u_{\mathrm{Suhl}} = 2.10 \times 10^6~\mathrm{J\,m^{-3}}$). (a)~Transfer rate $\Gamma_{\mathrm{transfer}}$ (left) and prompt thermal power $P_{\mathrm{th}}$ (right) versus FMR frequency $f_M$, following $\Gamma_{\mathrm{transfer}}(f_M) \propto f_M^{-10.82}[1 + (f_{\mathrm{crit}}/f_M)^6]^{-3.8}$ ($f_{\mathrm{crit}} = 0.62~\mathrm{GHz}$). Filled circles mark Ka, X ($9.5~\mathrm{GHz}$, $8.3 \times 10^4~\mathrm{s}^{-1}$, $0.32~\mu\mathrm{W}$), S ($2.45~\mathrm{GHz}$, $0.74~\mathrm{W}$), and L ($915~\mathrm{MHz}$, $31.7~\mathrm{kW}$) waveguide/resonator bands; dashed blue line marks a $1~\mathrm{kW}$ duty-averaged RF reference ($4.4\%$ of the $22.9~\mathrm{kW}$ peak Gilbert dissipation at $915~\mathrm{MHz}$, where instantaneous in-pulse $Q_{\mathrm{sci}} \approx 1.4$). (b)~In-plane Kittel bias field $B_{\mathrm{DC}}(f_M)$ (solid green) and single-layer FMR skin depth $\delta_{\mathrm{RF}}(f_M)$ (dashed purple).}
\label{fig:freq_scaling}
\end{figure}

Partitioning $\Delta M c^2 = \sum_{j=1}^m \epsilon_j$ across $m$ acceptor transitions yields a bare multi-transition Golden Rule rate $\Gamma_{\mathrm{transfer}}^{(0)}(E_{\mathrm{virt}}) = (2\pi/\hbar)|\Delta\mathcal{U}_{\mathrm{mag}}|^2 \tilde{\rho}_N(E_{\mathrm{virt}}, E_M)$, where $\tilde{\rho}_N(E_{\mathrm{virt}}, E_M) \approx \alpha(E_M)\exp[\beta(E_M)E_{\mathrm{virt}}]$ locally approximates the Bethe Fermi-gas multi-acceptor state density \cite{Bethe1936} regularized by the Pad\'{e}/L\"{o}wdin strong-coupling factor $F(g_j) = g_j / (1 + 2g_j)$ for $g_j = |\mathcal{V}_{\mathrm{Rec},j}|/\epsilon_j$ (appendix~\ref{app:lowdin}), softening the weak-coupling slope $\beta(0) \approx 0.995~\mathrm{MeV}^{-1}$ to $\beta(E_M) = 0.72~\mathrm{MeV}^{-1}$ at $E_{M,\mathrm{eff}} = 8.4~\mathrm{J}$. Under phase-locked FMR drive, the $k=0$ condensate maintains external oscillator phase coherence far exceeding passive spin-wave ring-down ($\tau_M = 2/(\gamma_e\Delta H) \approx 1.1~\mathrm{ns}$). Across the coherent transfer window $\tau = (\Gamma_{\mathrm{transfer}} + \gamma_{\mathrm{nuc}})^{-1} \approx \Gamma_{\mathrm{transfer}}^{-1}$ (above slow Korringa nuclear dephasing $\gamma_{\mathrm{nuc}}$), each acceptor transition $j$ with reduced M1 strength $O_{\mathrm{Rec},j} \equiv [B(\mathrm{M1}; j)\uparrow / \mu_N^2]^{1/2}$ exchanges magnons at single-channel rate $\Gamma_M(j)$, summing across the $\mathcal{N}_{\mathrm{M1}}$ accessible spin-flip states of the fragmented M1 doorway manifold $\mathcal{S}_{\mathrm{M1}}$ to yield the collective exchange rate $\Gamma_M^{\mathrm{eff}}$:
\begin{subequations}
\label{eq:gamma_m_all}
\begin{align}
\Gamma_M(j) &= \frac{2}{\hbar}\left( \mu_N O_{\mathrm{Rec},j} \right)\sqrt{\frac{2\mu_0 E_M}{V_{\mathrm{eff}}}}\sqrt{N_{\mathrm{Rec},j}}, \label{eq:gamma_m} \\
\Gamma_M^{\mathrm{eff}} &= \sum_{j \in \mathcal{S}_{\mathrm{M1}}} \Gamma_M(j) \equiv \mathcal{N}_{\mathrm{M1}}\bar{\Gamma}_M. \label{eq:gamma_m_eff}
\end{align}
\end{subequations}
Under the finite-time Lippmann--Schwinger propagator (appendix~\ref{app:diffusion}), collective magnon exchange across $n_j$ active partitions dresses \textit{intermediate off-shell} doorway states with a Gaussian interaction-energy fluctuation $f_{\epsilon_M}(\epsilon_M, \tau) = (2\pi\sigma_{\epsilon_M}^2)^{-1/2}\exp(-\epsilon_M^2 / (2\sigma_{\epsilon_M}^2))$ of root-mean-square width
\begin{equation}
\sigma_{\epsilon_M}(\tau) = \frac{1}{2}\left( \hbar\omega_M \right) n_j \Gamma_M^{\mathrm{eff}} \tau = \frac{\left( \hbar\omega_M \right) n_j \Gamma_M^{\mathrm{eff}}}{2\Gamma_{\mathrm{transfer}}}, \label{eq:sigma_m}
\end{equation}
while asymptotic ($t \gg \tau$) time evolution strictly enforces on-shell energy conservation $\sum_{j=1}^m \epsilon_j = \Delta M c^2 = 23.85~\mathrm{MeV}$ without extracting net MeV energy from the RF bath (appendix~\ref{app:diffusion}).

\begin{figure}[htbp]
\centering
\includegraphics[width=0.55\textwidth]{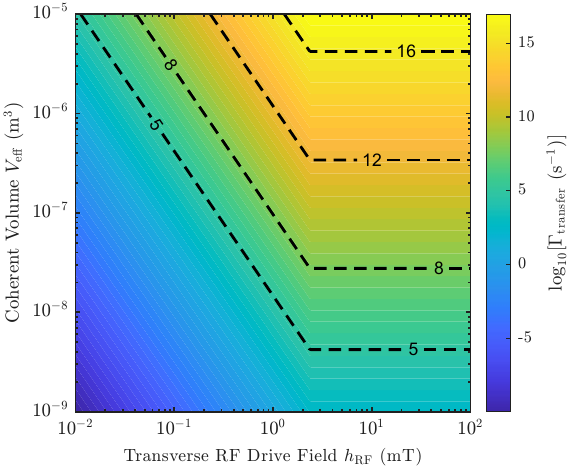}
\caption{Pre-asymptotic phase-space map of $\log_{10}[\Gamma_{\mathrm{transfer}}~(\mathrm{s}^{-1})]$ versus transverse RF field $h_{\mathrm{RF}}$ and coherent target volume $V_{\mathrm{eff}}$ ($n_{\mathrm{D}_2} = 1.89 \times 10^{27}~\mathrm{m}^{-3}$, $n_{\mathrm{Rec}} = 2.50 \times 10^{28}~\mathrm{m}^{-3}$). Below the Suhl threshold ($h_{\mathrm{RF}} < h_{\mathrm{crit}} = 2.3~\mathrm{mT}$), $\Gamma_{\mathrm{transfer}} \propto V_{\mathrm{eff}}^{3.67}h_{\mathrm{RF}}^{5.33}$; above $h_{\mathrm{crit}}$, parametric $\mathbf{k} \neq 0$ scattering clamps $h_{\mathrm{clamped}} = \min(h_{\mathrm{RF}}, h_{\mathrm{crit}})$, plateauing the dashed iso-rate contours and mandating volumetric scaling along $V_{\mathrm{eff}}$.}
\label{fig:suhl_phase}
\end{figure}

\begin{figure}[htbp]
\centering
\includegraphics[width=0.55\textwidth]{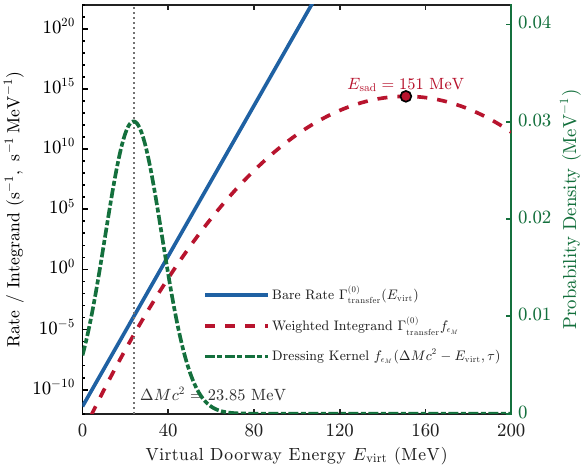}
\caption{Self-consistent Lippmann--Schwinger off-shell doorway convolution at $E_{M,\mathrm{eff}} = 8.4~\mathrm{J}$ ($f_M = 915~\mathrm{MHz}$, $|\Delta\mathcal{U}_{\mathrm{mag}}|^2 = 8.70 \times 10^{-11}~\mathrm{eV}^2$, $\beta = 0.72~\mathrm{MeV}^{-1}$, $\kappa = 6.76$), showing bare rate $\Gamma_{\mathrm{transfer}}^{(0)}(E_{\mathrm{virt}})$ (solid blue, left axis), Gaussian magnon dressing kernel $f_{\epsilon_M}(\Delta M c^2 - E_{\mathrm{virt}},\tau)$ centered at the on-shell energy $\Delta M c^2 = 23.85~\mathrm{MeV}$ with $\sigma_{\epsilon_M} = 13.3~\mathrm{MeV}$ (dash-dotted green, right axis), and weighted integrand $\Gamma_{\mathrm{transfer}}^{(0)}f_{\epsilon_M}$ peaking at the virtual saddle point $E_{\mathrm{sad}} = 151~\mathrm{MeV}$ and integrating to $\Gamma_{\mathrm{transfer}} = 8.30 \times 10^{15}~\mathrm{s}^{-1}$ (dashed red, left axis).}
\label{fig:quantum_diffusion}
\end{figure}

Unlike linear acoustic phonons, ferromagnetic spin waves obey the non-linear Landau--Lifshitz--Gilbert equation. Pumping the $k=0$ mode above Suhl's second-order instability threshold $h_{\mathrm{crit}}$ triggers parametric four-magnon decay ($\omega_0 + \omega_0 \to \omega_{\mathbf{k}} + \omega_{-\mathbf{k}}$) into spatially incoherent $\mathbf{k} \neq 0$ modes that cannot sustain Dicke supertransfer \cite{Suhl1957}:
\begin{subequations}
\label{eq:suhl_all}
\begin{align}
h_{\mathrm{crit}} &= \Delta H\sqrt{\frac{\Delta H}{\mu_0 M_s}}, \label{eq:hcrit} \\
E_{M,\mathrm{eff}} &= \frac{V_{\mathrm{eff}}}{2\mu_0}\mu_{r,\mathrm{eff}}^2\min\!\left( h_{\mathrm{RF}}^2, \; h_{\mathrm{crit}}^2 \right). \label{eq:em_eff}
\end{align}
\end{subequations}
Evaluating the off-shell doorway convolution at $E_M = E_{M,\mathrm{eff}}$ yields two distinct scaling regimes separated by a low-field rollover frequency $f_{\mathrm{crit}}$ (appendix~\ref{app:diffusion}):
\begin{subequations}
\label{eq:conv_all}
\begin{align}
\Gamma_{\mathrm{transfer}} &= \Gamma_{\mathrm{transfer}}^{(0)}(\Delta M c^2)\exp\!\left( \frac{1}{2}\beta^2\sigma_{\epsilon_M}^2 \right), \label{eq:conv_int} \\
\Gamma_{\mathrm{transfer}}^{\mathrm{pre}} &= \frac{2\pi}{\hbar}|\Delta\mathcal{U}_{\mathrm{mag}}|^2 \tilde{\rho}_{\mathrm{eff}}(\Delta M c^2, E_{M,\mathrm{eff}}) \propto V_{\mathrm{eff}}^{11/3}\min\!\left(h_{\mathrm{RF}}, h_{\mathrm{crit}}\right)^{16/3}, \label{eq:conv_pre} \\
\Gamma_{\mathrm{transfer}}^{\mathrm{sat}} &= \frac{\Gamma_M^{\mathrm{eff}}}{2\sqrt{2}}\frac{\hbar\omega_M}{\delta E - \beta\sigma_0^2}\left( \frac{\beta\Delta M c^2}{\kappa} + 2\kappa \right), \label{eq:conv_closed}
\end{align}
\end{subequations}
where $\kappa \equiv [\ln(\Gamma_{\mathrm{transfer}}/\Gamma_{\mathrm{transfer}}^{(0)})]^{1/2} = \beta\sigma_{\epsilon_M}/\sqrt{2}$, $n_j = (\Delta M c^2 + 2\kappa^2/\beta)/(\delta E - \beta\sigma_0^2)$, $\delta E \approx 6.75~\mathrm{MeV}$, and $\sigma_0 \approx 1.28~\mathrm{MeV}$ characterize the dipole strength centroid and spread \cite{Heyde2010,Kopecky1987,Muscher2020}. In the pre-asymptotic regime ($f_M > f_{\mathrm{crit}}$, equation~(\ref{eq:conv_pre})), progressive saturation of $F(g_j)$ across $m$-acceptor partitions yields $\tilde{\rho}_{\mathrm{eff}} \propto E_{M,\mathrm{eff}}^{5/3}$, producing the steep $\Gamma_{\mathrm{transfer}} \propto V_{\mathrm{eff}}^{3.67}h_{\mathrm{RF}}^{5.33}$ scaling below $h_{\mathrm{crit}}$ (figure~\ref{fig:suhl_phase}). Equation~(\ref{eq:conv_closed}) enforces self-consistency at the off-shell saddle point $E_{\mathrm{sad}} = \Delta M c^2 + 2\kappa^2/\beta$ (reaching $\sigma_{\epsilon_M} = 13.3~\mathrm{MeV}$, $\kappa = 6.76$, and $E_{\mathrm{sad}} = 151~\mathrm{MeV}$ at $915~\mathrm{MHz}$, figure~\ref{fig:quantum_diffusion}), where $e^{-G_{5\mathrm{D}}}$ enters only weakly via $\kappa$.

Equations~(\ref{eq:suhl_all}) and (\ref{eq:conv_all}) govern the scaling of a cryogenically cooled ($T_0 = 77\text{--}150~\mathrm{K}$) ferromagnetic $\mathrm{Pd}_{0.8}\mathrm{Ni}_{0.2}\mathrm{D}_x$ target. Unloaded $\mathrm{Pd}_{0.8}\mathrm{Ni}_{0.2}$ is an FCC ferromagnet ($0.6$ $d$-band holes per atom, Curie temperature $T_C \approx 230\text{--}260~\mathrm{K}$) \cite{Ododo1983} retaining high $\beta$-phase solubility ($\mathrm{D}/(\mathrm{Pd}+\mathrm{Ni}) \approx 0.8\text{--}0.9$) up to $20~\mathrm{at.\%~Ni}$ in thin films \cite{Grden1999,Hughes1992}. Operating at $T_0 \ll T_C$ in the vacancy-ordered $\mathrm{Pd}_3\vac\mathrm{D}_4$ phase ($n_{\mathrm{D}_2} \approx 1.89 \times 10^{27}~\mathrm{m}^{-3}$, $n_{\mathrm{Rec}} \approx 2.50 \times 10^{28}~\mathrm{m}^{-3}$) \cite{Fukai1994} -- or exchange-coupling sub-micron $\mathrm{Pd}_{0.8}\mathrm{Ni}_{0.2}\mathrm{D}_x$ layers to thin $\mathrm{Ni}$ pinning interlayers ($T_{C,\mathrm{Ni}} = 627~\mathrm{K}$) to counteract $1s\text{--}4d$ band-filling suppression -- sustains $\mu_0 M_s \approx 0.18~\mathrm{T}$, $\Delta H \approx 10~\mathrm{mT}$, and $\mu_{r,\mathrm{eff}} \approx 10^3$. Suhl clamping then fixes $B_{M,\mathrm{eff}} \approx 2.3~\mathrm{T}$, $u_{\mathrm{Suhl}} = E_{M,\mathrm{eff}}/V_{\mathrm{eff}} = 2.10 \times 10^6~\mathrm{J\,m^{-3}}$, and $U_{\mathrm{mag}} \approx 3.4 \times 10^{-27}~\mathrm{eV}$ ($U_e \approx 1.15~\mathrm{keV}$ \cite{Raiola2002,Karahadian2026,Cvetinovic2023}).

Because $u_{\mathrm{Suhl}}$ is clamped intensively (figure~\ref{fig:suhl_phase}), macroscopic scaling requires expanding $V_{\mathrm{eff}}(f_M)$ by lowering $f_M$ (figure~\ref{fig:freq_scaling}). Stacking $N_{\mathrm{lam}} \propto f_M^{-1/2}$ insulated ferromagnetic laminations of FMR skin depth $\delta_{\mathrm{RF}} = [2\rho_{\mathrm{el}} / (\omega_M \mu_0 \mu_\perp)]^{1/2} \propto f_M^{-1/2}$ ($\mu_\perp \approx \mu_0 M_s/\Delta H = 18$, expanding from $2.0~\mu\mathrm{m}$ at $9.5~\mathrm{GHz}$ to $6.44~\mu\mathrm{m}$ at $915~\mathrm{MHz}$, figure~\ref{fig:freq_scaling}(b)) inside a uniform-phase lumped-element loop-gap resonator \cite{Froncisz1982} of aperture $\propto \lambda_{\mathrm{RF}}^2 \propto f_M^{-2}$ yields $V_{\mathrm{eff}}(f_M) \propto f_M^{-3}$ and $E_{M,\mathrm{eff}}(f_M) = u_{\mathrm{Suhl}}V_{\mathrm{eff}}(f_M) \propto f_M^{-3}$. Substituting $V_{\mathrm{eff}} \propto f_M^{-3}$ into equation~(\ref{eq:conv_pre}) gives $\Gamma_{\mathrm{transfer}}(f_M) \propto f_M^{-10.8}$ above $f_{\mathrm{crit}} \approx 0.62~\mathrm{GHz}$ (figure~\ref{fig:freq_scaling}(a)), below which $B_{\mathrm{DC}}(f_{\mathrm{crit}}) \approx 2.7~\mathrm{mT}$ drops to the scale of $h_{\mathrm{crit}}$ and thin-film domain-saturation fields, triggering three-magnon and multi-domain dephasing alongside $F(g_j) \to \frac{1}{2}$ saturation in equation~(\ref{eq:conv_closed}). Whereas peak in-pulse Gilbert dissipation at $h_{\mathrm{crit}}$ scales linearly with volume as $P_{\mathrm{FMR,crit}} = [\pi f_M \Delta H^2 / (4\mu_0)]V_{\mathrm{eff}} \propto f_M^{-2}$ ($22.9~\mathrm{kW}$ peak in-pulse at $915~\mathrm{MHz}$, or $1~\mathrm{kW}$ time-averaged at $4.4\%$ pulse duty factor), in-pulse fusion power scales as $P_{\mathrm{th}} \propto V_{\mathrm{eff}}^{3.67} \propto f_M^{-10.8}$, spanning from diagnostic X-band rates ($9.5~\mathrm{GHz}$, $V_{\mathrm{eff}} = 4 \times 10^{-9}~\mathrm{m}^3$, $\Gamma_{\mathrm{transfer}} \approx 8.3 \times 10^4~\mathrm{s}^{-1}$) to net in-pulse scientific gain at L-band ($915~\mathrm{MHz}$, $V_{\mathrm{eff}} = 4 \times 10^{-6}~\mathrm{m}^3$, $\Gamma_{\mathrm{transfer}} \approx 8.3 \times 10^{15}~\mathrm{s}^{-1}$, $P_{\mathrm{th}} \approx 31.7~\mathrm{kW}$, instantaneous $Q_{\mathrm{sci}} = P_{\mathrm{th}}/P_{\mathrm{FMR,crit}} \approx 1.4$).

\section{Reaction kinematics and solid-state fusion anomalies}
\label{sec:kinematics}

\subsection{Multi-acceptor de-excitation versus single-acceptor breakup signatures}
\label{subsec:breakup}

Asymptotic reaction products depend sharply on the on-shell acceptor multiplicity $m$. Partitioning $\Delta M c^2 = 23.85~\mathrm{MeV}$ across $m = 3\text{--}8$ host nuclei ($\epsilon_j \approx 3.0\text{--}8.0~\mathrm{MeV}$) populates bound $1^+$ levels below host nucleon separation thresholds ($S_n = 7.03\text{--}10.47~\mathrm{MeV}$ across $^{102\text{--}110}\mathrm{Pd}$; $S_n = 12.22~\mathrm{MeV}$ in $^{58}\mathrm{Ni}$), strictly forbidding neutron or charged-particle emission. In the strong-exchange hypothesis ($\Gamma_M^{\mathrm{eff}} \gg \Gamma_\gamma \sim 10^{10}\text{--}10^{13}~\mathrm{s}^{-1}$), rapid coherent magnon exchange coupled to anharmonic magnetoelastic damping down-converts these sub-threshold excitations radiationlessly into the magnon--phonon bath; conversely, if micro-eV sideband dressing does not complete secular down-conversion within $\Gamma_\gamma^{-1}$, $m = 3\text{--}8$ events will instead emit a characteristic prompt $3\text{--}9~\mathrm{MeV}$ M1 spin-flip $\gamma$-ray cascade synchronized with FMR resonance (section~\ref{sec:limitations}).

In rare single-acceptor events ($m = 1$, combinatorially suppressed by $\tilde{\rho}_N^{(m=1)}/\tilde{\rho}_N \sim 10^{-10}\text{--}10^{-7}$; figure~\ref{fig:energy_pathway}, Zone~III), a single $^{106}\mathrm{Pd}$ or $^{58}\mathrm{Ni}$ nucleus absorbs the full on-shell $E^* = 23.85~\mathrm{MeV}$ quantum via giant-resonance tails with small $O_{\mathrm{Rec},j}$ that escape coherent down-conversion. Alongside Hauser--Feshbach evaporation ($T_{\mathrm{evap}} \approx 1.5\text{--}2.5~\mathrm{MeV}$), pre-equilibrium two-body breakup to daughter ground states ($Q_0 = E^* - S_{n,\alpha}$) produces monoenergetic kinematic endpoints for $^{106}\mathrm{Pd}$ ($S_n = 9.56~\mathrm{MeV}$, $S_\alpha = 3.20~\mathrm{MeV}$) and $^{58}\mathrm{Ni}$ ($S_n = 12.22~\mathrm{MeV}$):
\begin{subequations}
\label{eq:breakup_all}
\begin{align}
{}^{106}\mathrm{Pd}^* &\to {}^{105}\mathrm{Pd} + n_0: \qquad E_n \approx \frac{105}{106}Q_0 = 14.15~\mathrm{MeV}, \label{eq:pd_n} \\
{}^{58}\mathrm{Ni}^* &\to {}^{57}\mathrm{Ni} + n_0: \qquad E_n \approx \frac{57}{58}Q_0 = 11.43~\mathrm{MeV}, \label{eq:ni_n} \\
{}^{106}\mathrm{Pd}^* &\to {}^{102}\mathrm{Ru} + \alpha_0: \qquad E_\alpha \approx \frac{102}{106}Q_0 = 19.87~\mathrm{MeV}. \label{eq:pd_alpha}
\end{align}
\end{subequations}
Because $Q_0 = 20.65~\mathrm{MeV}$ exceeds the $\alpha + {}^{102}\mathrm{Ru}$ Coulomb barrier ($V_C \approx 14.5~\mathrm{MeV}$) and the $\sim 95~\mu\mathrm{m}$ stopping range of $19.87~\mathrm{MeV}$ $\alpha$ particles in $\mathrm{Pd}$ far exceeds $\delta_{\mathrm{RF}} = 2.0\text{--}6.4~\mu\mathrm{m}$, prompt $\alpha_0$ emission provides a background-free thin-film spectroscopic diagnostic. Conventional $2.45~\mathrm{MeV}$ neutrons and $3.02~\mathrm{MeV}$ protons arise solely from trace tunnel decay of the open $|{}^4\mathrm{He}^*(3+1)\rangle$ doorway resonance (appendix~\ref{app:lowdin}) \cite{Tilley1992}:
\begin{equation}
\frac{\Gamma_{3+1}}{\Gamma_{\mathrm{transfer}}} = \frac{1}{2}\frac{(\hbar\Gamma_{\mathrm{transfer}})(\hbar\gamma_{\mathrm{tunnel}})}{|\mathcal{V}_{\mathrm{mag}}|^2 + \hbar^2\gamma_{\mathrm{tunnel}}^2} \approx \frac{\hbar\Gamma_{\mathrm{transfer}}}{2\hbar\gamma_{\mathrm{tunnel}}} \ll 1, \label{eq:br_3p1}
\end{equation}
where $|\mathcal{V}_{\mathrm{mag}}|^2 \equiv U_{\mathrm{mag}}^2 N_{\mathrm{D}_2}N_{\mathrm{Rec}} \ll \hbar^2\gamma_{\mathrm{tunnel}}^2$, giving $\Gamma_{3+1}/\Gamma_{\mathrm{transfer}} \approx 4.6 \times 10^{-18}$ at $9.5~\mathrm{GHz}$ and $4.6 \times 10^{-7}$ at $915~\mathrm{MHz}$ with $n/p$ branching $R \approx 0.92\text{--}0.93$ \cite{Huke2008,Tilley1992}.

\subsection{Reconciliation of solid-state nuclear anomalies}
\label{subsec:anomalies}

This framework unifies five persistent experimental observations across the solid-state nuclear physics literature:
\begin{enumerate}
\item \textit{Periodic-table spin-susceptibility bifurcation:} Large screening potentials in partially filled $d$-band ferromagnets ($\mathrm{Fe}, \mathrm{Co}, \mathrm{Ni}$: $200\text{--}450~\mathrm{eV}$) and Stoner paramagnets ($\mathrm{Pd}, \mathrm{Rh}, \mathrm{Pt}$: $440\text{--}840~\mathrm{eV}$) versus gas-like values in filled-$d$-shell noble metals ($\mathrm{Cu}, \mathrm{Ag}, \mathrm{Au}$: $23\text{--}61~\mathrm{eV}$) \cite{Kasagi2002,Raiola2002,Huke2008} mirror both $m^*$-enhanced static screening \cite{Czerski2016} and the Stoner $s\text{--}d$ hyperfine susceptibility $\mu_{r,\mathrm{eff}} \propto \chi_d$, which collapses once the $d$ shell fills.
\item \textit{Passive thermal baseline versus active FMR-RET:} In unstimulated $\mathrm{Pd}$ foils ($\mathbf{B}_{\mathrm{DC}} = \mathbf{h}_{\mathrm{RF}} = 0$), Lipoglav\v{s}ek and Cvetinovi\'{c} \cite{Lipoglavsek2026} observed slow incoherent $3.0~\mathrm{MeV}$ proton emission ($\sim 10^{-2}~\mathrm{s}^{-1}$) from defect-screened sites ($U_e = 475~\mathrm{eV}$) with zero $0^+$ $e^-e^+$ pairs. Because $0^+ \to 0^+$ single-photon decay is forbidden and pair decay is absent, accessing the $23.85~\mathrm{MeV}$ ground state requires transferring $\Delta J = 1$ angular momentum to an actively pumped magnon bath ($E_{M,\mathrm{eff}} > 0$).
\item \textit{Sub-keV yield plateau:} Once sub-keV beams and electron-cyclotron-resonance microwaves populate near-surface $\mathrm{Pd}_3\vac\mathrm{D}_4$ traps \cite{Karahadian2026}, $\Gamma_{\mathrm{transfer}}$ (equation~(\ref{eq:conv_closed})) and its $3+1$ proton leakage (equation~(\ref{eq:br_3p1})) depend on trapped inventory $N_{\mathrm{D}_2}$ and $\kappa$ rather than projectile energy $E_{\mathrm{cm}}$.
\item \textit{Energetic charged particles and fast neutrons:} Anomalous $5\text{--}28~\mathrm{MeV}$ charged particles \cite{Chambers1990,Chambers1991} and $>2.45~\mathrm{MeV}$ neutrons \cite{Takahashi1990} in tritium-free deuterides match our $m=1$ breakup endpoints (equation~(\ref{eq:breakup_all})).
\item \textit{Magnetic and RF gating:} Empirical magnetic and RF triggering reflects opening of the FMR M1 channel alongside ferromagnetic Overhauser lifting of spin cancellation \cite{Overhauser1953} in the secondary Dirac pathway (appendix~\ref{app:dirac}).
\end{enumerate}

\section{Experimental regimes and model limitations}
\label{sec:limitations}

Translating the Magnon-RET scaling law into a controlled laboratory benchmark imposes three experimental boundary conditions and defines sharp falsification tests:
\begin{enumerate}
\item \textit{Cryogenic pulsed thermal management and ferromagnetic stability:} At the Suhl clamping limit ($h_{\mathrm{RF}} = h_{\mathrm{crit}} = 2.3~\mathrm{mT}$), Gilbert spin-lattice damping deposits $p_{\mathrm{FMR,crit}} = \pi f_M \Delta H^2 / (4\mu_0) \approx 5.7~\mathrm{kW\,cm^{-3}}$ at $915~\mathrm{MHz}$, which under continuous-wave drive would heat an uncooled foil across $T_C \approx 230\text{--}260~\mathrm{K}$ \cite{Ododo1983} and trigger $\mathrm{D}_2$ desorption within milliseconds. Sustaining ferromagnetic order and $\mathrm{Pd}_3\vac\mathrm{D}_4$ stoichiometry therefore requires liquid-nitrogen cryogenic cooling ($T_0 = 77\text{--}120~\mathrm{K} \ll T_C$), low-duty-cycle microsecond RF pulsing ($\tau_{\mathrm{pulse}} \sim 1\text{--}50~\mu\mathrm{s} \gg \tau_M, \tau$), and exchange-coupled $\mathrm{Ni}/\mathrm{Pd}_{0.8}\mathrm{Ni}_{0.2}\mathrm{D}_x$ multilayer wafer stacks in which $5\text{--}15~\mathrm{nm}$ $\mathrm{Ni}$ interlayers pin $M_s$ against hydride $d$-band filling while narrowing $\Delta H$ (since $P_{\mathrm{FMR,crit}} \propto \Delta H^2$ whereas $h_{\mathrm{crit}} \propto \Delta H^{3/2}$).
\item \textit{Spatial Dicke phase uniformity:} Macroscopic supertransfer across $V_{\mathrm{eff}} \sim \mathrm{cm}^3$ requires uniform RF magnetic phase across the laminated target stack so that $\sum_k \langle \hat{\mathbf{B}}_{M,k} \rangle$ adds constructively without standing-wave sign reversals. Sub-wavelength dielectric-loaded loop-gap resonators \cite{Froncisz1982} isolate the electric field inside high-permittivity capacitive gaps while maintaining a spatially uniform axial RF magnetic mode $\mathbf{h}_{\mathrm{RF}}$ across the inductive bore.
\item \textit{Falsifiable radiative versus non-radiative signatures:} While our off-shell Lippmann--Schwinger treatment models secular down-conversion of $m = 3\text{--}8$ sub-threshold excitations phenomenologically via $\Gamma_M^{\mathrm{eff}} \gg \Gamma_\gamma$, a microscopic non-perturbative treatment of MeV-to-meV boson cascading remains an open theoretical problem. Crucially, the Magnon-RET mechanism is experimentally falsifiable independent of whether sub-threshold thermalization is complete: sweeping $B_{\mathrm{DC}}$ through the Kittel resonance $B_{\mathrm{DC}}^{\mathrm{FMR}}(f_M)$ at fixed RF input power must produce a resonant, Suhl-saturating step in (i)~$19.87~\mathrm{MeV}$ $\alpha_0$ particles on thin-film charged-particle detectors, (ii)~$14.15$ and $11.43~\mathrm{MeV}$ $n_0$ breakup neutrons, and (iii)~prompt $3\text{--}9~\mathrm{MeV}$ M1 $\gamma$ rays in the event of incomplete Zeno down-conversion, all vanishing identically in $\mathrm{H}_2$-loaded control wafers or off-resonant $B_{\mathrm{DC}}$ detuning.
\end{enumerate}

\section{Conclusion}
\label{sec:conclusion}

Extending the collective nuclear Dicke model to FMR-driven coherent magnons eliminates inertial rest-mass dilution via Turov--Petrov hyperfine-amplified M1 coupling to host $0^+ \to 1^+$ spin-flip giant resonances. Enforcing Suhl spin-wave clamping and Feshbach channel-closing doorway asymmetry in the off-shell Lippmann--Schwinger dressing integral yields a pre-asymptotic volumetric scaling law ($\Gamma_{\mathrm{transfer}} \propto f_M^{-10.8}$) under cryogenic pulsed FMR operation, predicting falsifiable $14.15~\mathrm{MeV}$ ($^{106}\mathrm{Pd}$) and $11.43~\mathrm{MeV}$ ($^{58}\mathrm{Ni}$) $n_0$ neutrons, $19.87~\mathrm{MeV}$ $\alpha_0$ particles, and FMR-gated M1 $\gamma$/calorimetric signatures alongside $10^{-7}$-suppressed $2.45~\mathrm{MeV}$ neutron leakage.

\appendix
\numberwithin{equation}{section}
\makeatletter
\renewcommand{\@seccntformat}[1]{Appendix \csname the#1\endcsname.\quad}
\makeatother

\section{Second-order L\"{o}wdin--Feshbach resolvent and channel-closing $|1-\eta|$ doorway asymmetry}
\label{app:lowdin}

To derive equations~(\ref{eq:dicke_u}) and (\ref{eq:br_3p1}) with exact composite-subsystem energy accounting, we partition the magnon-nuclear Hilbert space ($\hat{H}_0 = \hat{H}_{A=4} + \hat{H}_{\mathrm{Rec}} + \hat{H}_M$) via orthogonal Feshbach--L\"{o}wdin projectors $\hat{P} + \hat{Q} + \hat{D} = \hat{\mathbb{I}}$ \cite{Lowdin1951,Lowdin1962,Feshbach1958,Feshbach1962,Feshbach1967}. Here $\hat{P}$ projects onto the resonant on-shell manifold near $E_i = \Delta M c^2 + n_M\hbar\omega_M$, containing the initial collective donor Dicke state $|i\rangle = |\mathrm{D}_2^{\mathrm{ens}}; 0_{\mathrm{Rec}}; n_M\rangle$ ($\Delta E_{\mathrm{Rec}} = 0$, $E_{A=4} = \Delta M c^2 = 23.85~\mathrm{MeV}$) and final $m$-acceptor states $|f\rangle = |{}^4\mathrm{He}; \{1_j\}_{j=1}^m; n_M'\rangle$ ($E_{A=4} = 0$, $\sum_{j=1}^m \epsilon_j \approx \Delta M c^2$); $\hat{Q}$ projects onto off-resonant single-vertex intermediate states coupled to $\hat{P}$ by $\hat{H}_{\mathrm{int}} = -\sum_k \bm{\mu}_{N,k}\cdot\hat{\mathbf{B}}_{M,k}$; and $\hat{D}$ projects onto composite doorway configurations $|d; \Phi_{\mathrm{Rec}}\rangle$ built on the open $A=4$ resonance $|d\rangle = |{}^4\mathrm{He}^*(3+1)\rangle$ ($E_d = 24.25~\mathrm{MeV}$ above $|{}^4\mathrm{He}\rangle$, Gamow width $\hbar\gamma_{\mathrm{tunnel}} \approx 5.95~\mathrm{MeV}$ above the $19.81~\mathrm{MeV}$ $p+t$ and $20.58~\mathrm{MeV}$ $n+{}^3\mathrm{He}$ thresholds) \cite{Tilley1992,Feshbach1967}.

Eliminating $\hat{D}$ and $\hat{Q}$ via L\"{o}wdin partitioning gives the effective Hamiltonian on $\hat{P}$:
\begin{subequations}
\label{eq:lowdin_all}
\begin{align}
\hat{H}_{\mathrm{eff}}(E) &= \hat{P}\hat{\mathcal{H}}_P(E)\hat{P} + \hat{P}\tilde{H}_{\mathrm{int}}(E)\hat{Q}\frac{1}{E - \hat{Q}\hat{\mathcal{H}}_Q(E)\hat{Q}}\hat{Q}\tilde{H}_{\mathrm{int}}(E)\hat{P}, \label{eq:lowdin_eff} \\
\begin{split}
\hat{\mathcal{H}}_{P,Q}(E) &= \hat{H}_0 + \hat{V}_{P,Q;D}\Bigl[E - \hat{H}_{\mathrm{Rec}} - \hat{H}_{DD} \\
&\quad + \tfrac{i}{2}\hbar\hat{\gamma}_D(E - \hat{H}_{\mathrm{Rec}})\Bigr]^{-1}\hat{V}_{D;P,Q},
\end{split} \label{eq:hq_def}
\end{align}
\end{subequations}
where $\hbar\hat{\gamma}_D(E_{A=4}^{\mathrm{sub}})$ is the energy-dependent $3+1$ continuum decay width evaluated at the available $A=4$ subsystem energy $E_{A=4}^{\mathrm{sub}} = E - \Delta E_{\mathrm{Rec}}$, which equals $\hbar\gamma_{\mathrm{tunnel}} \approx 5.95~\mathrm{MeV}$ for $E_{A=4}^{\mathrm{sub}} \approx 23.85~\mathrm{MeV}$ but vanishes identically ($\hat{\gamma}_D = 0$) below the $19.81~\mathrm{MeV}$ breakup threshold \cite{Feshbach1958,Feshbach1967}, and $\tilde{H}_{\mathrm{int}}(E)$ includes the doorway-assisted M1 vertex. Two time-ordered pathways connect $|i\rangle$ to $|f\rangle$ at $E = E_i$:
\begin{enumerate}
\item \textit{Donor-first (lower) pathway ($\vert{}q_-\rangle = \vert{}{}^4\mathrm{He}~0^+; 0_{\mathrm{Rec}}; n_M \pm 1\rangle$):} Prior to acceptor excitation ($\Delta E_{\mathrm{Rec}} = 0$), $|i\rangle$ mixes ($W \approx 1.15~\mathrm{MeV}$) with $|d; 0_{\mathrm{Rec}}\rangle \in \hat{D}$ at above-threshold subsystem energy $E_{A=4}^{\mathrm{sub}} = 23.85~\mathrm{MeV}$, encountering the near-resonant open-channel doorway denominator $E_i - E_d + i\hbar\gamma_{\mathrm{tunnel}}/2 = -0.40 + 2.98i~\mathrm{MeV}$ before de-exciting to $|q_-\rangle$ (with real $\hat{Q}$-resolvent denominator $E_i - E_{q_-} = +\Delta M c^2$).
\item \textit{Acceptor-first (upper) pathway ($\vert{}q_+\rangle = \vert{}\mathrm{D}_2^{\mathrm{ens}}; \Psi_{\mathrm{Rec}}; n_M \pm 1\rangle$):} Exciting the heavy-acceptor manifold first to $\Delta E_{\mathrm{Rec}} \approx \Delta M c^2 = 23.85~\mathrm{MeV}$ leaves $A=4$ subsystem energy $E_{A=4}^{\mathrm{sub}} = E_i - \Delta E_{\mathrm{Rec}} \approx 0~\mathrm{MeV}$ ($19.81~\mathrm{MeV}$ below the $p+t$ breakup threshold). Inside $|q_+\rangle$, virtual coupling to $|d; \Psi_{\mathrm{Rec}}\rangle \in \hat{D}$ is therefore kinematically closed ($\hat{\gamma}_D(0) = 0$) and off-resonant by $(E_i - \Delta E_{\mathrm{Rec}}) - E_d \approx -24.25~\mathrm{MeV}$, producing only a negligible real shift $\Delta_d(0) \approx -W^2 / E_d \approx -0.05~\mathrm{MeV}$ around $E_i - E_{q_+} \approx -\Delta M c^2$.
\end{enumerate}

Summing both time orderings in equation~(\ref{eq:lowdin_eff}) at $E = E_i$ yields the net effective coupling amplitude proportional to
\begin{equation}
\mathcal{G}_{\mathrm{net}} = \frac{\mathcal{A}_-(E_{A=4}^{\mathrm{sub}} = \Delta M c^2)}{\Delta M c^2} - \frac{\mathcal{A}_+(E_{A=4}^{\mathrm{sub}} \approx 0)}{\Delta M c^2 - \Delta_d(0)} \equiv \frac{\mathcal{A}_0(1 - \eta)}{\Delta M c^2}, \label{eq:eta_def}
\end{equation}
where $\mathcal{A}_0$ is the bare donor M1 capture amplitude and $\eta$ is the pathway ratio. In a closed Hermitian subspace without threshold channel closing ($\hat{\gamma}_D \to 0$, $\mathcal{A}_+ = \mathcal{A}_-$, $\Delta_d \to 0$), $\eta \to 1$ and the two pathways cancel identically. With open-channel doorway dressing active above threshold in $\mathcal{A}_-$ but suppressed by $-E_d$ below threshold in $\mathcal{A}_+$, destructive interference is broken at the level
\begin{equation}
\begin{aligned}
|1 - \eta| &\approx \left| \frac{W\,\chi_{di}}{\Delta M c^2 - E_d + i\hbar\gamma_{\mathrm{tunnel}}/2} \right| \\
&= \frac{W\,|\chi_{di}|}{\sqrt{(\Delta M c^2 - E_d)^2 + (\hbar\gamma_{\mathrm{tunnel}}/2)^2}} \approx 0.10,
\end{aligned}
\label{eq:eta_val}
\end{equation}
for reduced doorway-to-donor capture ratio $|\chi_{di}| \approx 0.26$.

In the strong-coupling regime where individual collective acceptor couplings $g_j \equiv |\mathcal{V}_{\mathrm{Rec},j}|/\epsilon_j$ approach or exceed unity, resumming diagonal and off-diagonal self-energy loops in $\hat{Q}\hat{\mathcal{H}}_Q\hat{Q}$ replaces the bare perturbative ratio $g_j$ at each acceptor vertex by the bounded Pad\'{e}/L\"{o}wdin strong-coupling regulator $F(g_j) = g_j / (1 + 2g_j) \le \frac{1}{2}$. Factoring the $N_{\mathrm{Rec}}$ phased-array sum out of the regularized $m$-fold state density
\begin{equation}
\tilde{\rho}_N(E_{\mathrm{virt}}, E_M) = \frac{1}{N_{\mathrm{Rec}}}\sum_f \left[\prod_{j=1}^m F^2(g_j)\right]\delta\!\left(E_{\mathrm{virt}} - \sum_{j=1}^m \epsilon_j\right) \label{eq:rho_reg}
\end{equation}
isolates $|\Delta\mathcal{U}_{\mathrm{mag}}|^2 = |1-\eta|^2 U_{\mathrm{mag}}^2 N_{\mathrm{D}_2}N_{\mathrm{Rec}}$ in equation~(\ref{eq:dicke_u}). Simultaneously, optical-theorem flux loss from $-\operatorname{Im}\hat{\mathcal{H}}_P(E_i)$ through the open $|d; 0_{\mathrm{Rec}}\rangle$ doorway during the transfer window $\tau = \Gamma_{\mathrm{transfer}}^{-1}$ yields the competing $3+1$ leakage ratio in equation~(\ref{eq:br_3p1}).

\section{Finite-time Lippmann--Schwinger off-shell dressing and secular energy conservation}
\label{app:diffusion}

To derive equations~(\ref{eq:sigma_m}) and (\ref{eq:conv_all}) while preserving strict asymptotic energy conservation, we evaluate the transition rate in the interaction picture generated by the fast collective magnon-acceptor exchange $\hat{H}_{\mathrm{exch}}(t)$ acting on intermediate doorway configurations alongside the second-order transfer vertex $\hat{V}_{\mathrm{RET}}$ ($\langle f | \hat{V}_{\mathrm{RET}} | i \rangle = \Delta\mathcal{U}_{\mathrm{mag}}\prod_{j=1}^m F(g_j)$). Across the finite coherent transfer lifetime $\tau = \Gamma_{\mathrm{transfer}}^{-1}$, the Lippmann--Schwinger transition rate summed over asymptotic final states $f$ is
\begin{equation}
\begin{aligned}
\Gamma_{\mathrm{transfer}} &= \lim_{T_{\mathrm{obs}} \to \infty} \frac{1}{\hbar^2 T_{\mathrm{obs}}}\sum_f \left| \int_0^{T_{\mathrm{obs}}} \mathrm{d}t \, e^{i(E_i - E_f)t/\hbar - t/(2\tau)} \right. \\
&\quad \times \left. \vphantom{\int_0^{T_{\mathrm{obs}}}} \langle f | \hat{U}_{\mathrm{exch}}^\dagger(t,0)\hat{V}_{\mathrm{RET}}\hat{U}_{\mathrm{exch}}(t,0) | i \rangle \right|^2,
\end{aligned}
\label{eq:ls_int}
\end{equation}
where $\hat{U}_{\mathrm{exch}}(t,0) = \mathcal{T}\exp[-(i/\hbar)\int_0^t \hat{H}_{\mathrm{exch}}(t')\,\mathrm{d}t']$. Crucially, the outer secular integral over $T_{\mathrm{obs}} \gg \tau \gg \hbar/\Delta M c^2$ enforces exact on-shell conservation of total asymptotic energy $E_f = E_i$, fixing the final nuclear excitation at $\sum_{j=1}^m \epsilon_j = \Delta M c^2 \pm \Delta N_M\hbar\omega_M = 23.85~\mathrm{MeV}$ (with net bath exchange $|\Delta N_M|\hbar\omega_M \ll \mathrm{eV}$, drawing negligible net energy from the RF drive). Within the transient window $\tau$, however, coherent magnon exchange across the $n_j$ active partitions of the intermediate doorway manifold at collective rate $\Gamma_M^{\mathrm{eff}}$ (equation~(\ref{eq:gamma_m_eff})) generates a ballistic quantum walk over $\Delta n_{M,\mathrm{rms}} \simeq \frac{1}{2}n_j \Gamma_M^{\mathrm{eff}} \tau$ virtual sidebands, dressing intermediate off-shell doorway states at $E_{\mathrm{virt}} = \Delta M c^2 - \epsilon_M$ with the normalized Gaussian spectral kernel
\begin{equation}
f_{\epsilon_M}(\epsilon_M, \tau) = \frac{1}{\sqrt{2\pi\sigma_{\epsilon_M}^2(\tau)}}\exp\!\left( -\frac{\epsilon_M^2}{2\sigma_{\epsilon_M}^2(\tau)} \right), \label{eq:gauss_b}
\end{equation}
where $\sigma_{\epsilon_M}(\tau) = \frac{1}{2}(\hbar\omega_M)n_j \Gamma_M^{\mathrm{eff}} \tau$ establishes equation~(\ref{eq:sigma_m}).

Integrating the intermediate doorway density $\tilde{\rho}_N(E_{\mathrm{virt}}, E_M) = \tilde{\rho}_N(\Delta M c^2, E_M)e^{-\beta\epsilon_M}$ against $f_{\epsilon_M}(\epsilon_M, \tau)$ and completing the square yields
\begin{equation}
\begin{aligned}
\Gamma_{\mathrm{transfer}} &= \frac{2\pi}{\hbar}|\Delta\mathcal{U}_{\mathrm{mag}}|^2 \tilde{\rho}_N(\Delta M c^2, E_M) \int_{-\infty}^{\infty} e^{-\beta\epsilon_M} f_{\epsilon_M}(\epsilon_M, \tau)\,\mathrm{d}\epsilon_M \\
&= \Gamma_{\mathrm{transfer}}^{(0)}(\Delta M c^2)\exp\!\left( \frac{1}{2}\beta^2\sigma_{\epsilon_M}^2 \right),
\end{aligned}
\label{eq:conv_b}
\end{equation}
recovering equation~(\ref{eq:conv_int}). In the pre-asymptotic regime ($f_M > f_{\mathrm{crit}}$), where $g_j \propto E_{M,\mathrm{eff}}^{1/2}$ transitions across the Pad\'{e} knee $F(g_j) = g_j/(1 + 2g_j)$ over dominant $m = 3\text{--}8$ partitions, numerical evaluation of equation~(\ref{eq:conv_b}) yields an effective dressed state density $\tilde{\rho}_{\mathrm{eff}}(\Delta M c^2, E_{M,\mathrm{eff}}) \equiv \tilde{\rho}_N(\Delta M c^2, E_{M,\mathrm{eff}})\exp(\frac{1}{2}\beta^2\sigma_{\epsilon_M}^2) \propto E_{M,\mathrm{eff}}^{5/3}$. Combining $|\Delta\mathcal{U}_{\mathrm{mag}}|^2 \propto V_{\mathrm{eff}}^2\min(h_{\mathrm{RF}}, h_{\mathrm{crit}})^2$ with $E_{M,\mathrm{eff}}^{5/3} \propto V_{\mathrm{eff}}^{5/3}\min(h_{\mathrm{RF}}, h_{\mathrm{crit}})^{10/3}$ gives $\Gamma_{\mathrm{transfer}}^{\mathrm{pre}} \propto V_{\mathrm{eff}}^{11/3}\min(h_{\mathrm{RF}}, h_{\mathrm{crit}})^{16/3}$, establishing equation~(\ref{eq:conv_pre}).

Defining $\kappa \equiv [\ln(\Gamma_{\mathrm{transfer}}/\Gamma_{\mathrm{transfer}}^{(0)})]^{1/2} = \beta\sigma_{\epsilon_M}/\sqrt{2}$, the virtual doorway integrand in equation~(\ref{eq:conv_b}) peaks at the off-shell saddle point $\epsilon_M^* = -\beta\sigma_{\epsilon_M}^2 = -2\kappa^2/\beta$, corresponding to virtual doorway energy $E_{\mathrm{sad}} = \Delta M c^2 - \epsilon_M^* = \Delta M c^2 + 2\kappa^2/\beta$ (above which both the Gaussian kernel and the Bethe Fermi-gas level density $\ln\tilde{\rho}_N(E_{\mathrm{virt}}) \propto 2\sqrt{a E_{\mathrm{virt}}}$ \cite{Bethe1936} bend the integrand downward, ensuring convergence). For an M1 spin-flip strength distribution of centroid $\delta E \approx 6.75~\mathrm{MeV}$ and width $\sigma_0 \approx 1.28~\mathrm{MeV}$ \cite{Heyde2010,Kopecky1987,Muscher2020}, exponential weighting shifts the effective energy per partition step to $\delta E_{\mathrm{eff}} = \delta E - \beta\sigma_0^2$, giving $n_j = (\Delta M c^2 + 2\kappa^2/\beta)/(\delta E - \beta\sigma_0^2)$. Substituting $\sigma_{\epsilon_M} = \sqrt{2}\kappa/\beta$, $\tau = \Gamma_{\mathrm{transfer}}^{-1}$, and $n_j$ into equation~(\ref{eq:sigma_m}) and solving for $\Gamma_{\mathrm{transfer}}$ gives the diffusion-saturated limit
\begin{equation}
\begin{aligned}
\Gamma_{\mathrm{transfer}}^{\mathrm{sat}} &= \frac{\beta(\hbar\omega_M)\Gamma_M^{\mathrm{eff}}}{2\sqrt{2}\kappa}\frac{\Delta M c^2 + 2\kappa^2/\beta}{\delta E - \beta\sigma_0^2} \\
&= \frac{\Gamma_M^{\mathrm{eff}}}{2\sqrt{2}}\frac{\hbar\omega_M}{\delta E - \beta\sigma_0^2}\left( \frac{\beta\Delta M c^2}{\kappa} + 2\kappa \right),
\end{aligned}
\label{eq:closed_b}
\end{equation}
establishing equation~(\ref{eq:conv_closed}).
\section{Secondary relativistic Dirac spin-dipole pathway and ferromagnetic Overhauser polarization}
\label{app:dirac}

Alongside the primary hyperfine M1 interaction (equation~(\ref{eq:magnon_ham})), FMR pumping also lifts thermal spin cancellation in the secondary relativistic center-of-mass boost channel (equation~(\ref{eq:phonon_int})) driven by magnetoelastic lattice coupling. Because the internal operator $\hat{\bm{\mathcal{O}}}_{\mathrm{boost}} = \sum_{n=1}^A \bm{\sigma}_n \times (\hat{\bm{\pi}}_n / m_n c)$ in equation~(\ref{eq:phonon_int}) is the cross product of an axial spin vector ($1^+$) and a polar internal momentum vector ($1^-$), it transforms as a rank-1 odd-parity ($K^\pi = 1^-$) relativistic spin-dipole operator ($\Delta J = 0, \pm 1$ with $0 \not\to 0$; $\Delta\pi = \text{yes}$), coupling rest-frame deuteron $^3\mathrm{S}_1/{}^3\mathrm{D}_1$ states to internal $^1\mathrm{P}_1$ continuum excitations ($E_2 - E_1 \approx 125.4~\mathrm{MeV}$) and $|\mathrm{D}_2~{}^3\mathrm{P}_1, M_J\rangle$ ($J^\pi = 1^-$) donor states to $|{}^4\mathrm{He}~0^+\rangle$ \cite{Hagelstein2023}.

By the Wigner--Eckart theorem, for quantization axis $\hat{\mathbf{z}} \parallel \mathbf{B}_{\mathrm{DC}}$ and spin-1 projections $M_J \in \{-1, 0, +1\}$, the off-diagonal deuteron boost matrix element satisfies \cite{Hagelstein2023}
\begin{equation}
H_{12}^{\mathrm{Dirac}} = \langle \Phi_{1\mathrm{P}} | \mathbf{a}\cdot c\hat{\mathbf{P}}_j | \Phi_{3\mathrm{S}/3\mathrm{D}}, M_J \rangle = \zeta_{\mathrm{D}} M_J c\hat{P}_{j,z}, \label{eq:dirac_mj}
\end{equation}
with $\zeta_{\mathrm{D}} \approx 2.98 \times 10^{-3}$. Summing coherently over $N_{\mathrm{D}_2}$ donor sites yields a collective Dicke amplitude proportional to $\sum_{k=1}^{N_{\mathrm{D}_2}} M_{J,k} = N_{\mathrm{D}_2}\langle M_J \rangle$, which vanishes identically ($\langle M_J \rangle = 0$) in an unpolarized thermal ensemble.

In a cryogenically cooled ferromagnetic $\mathrm{Pd}_{0.8}\mathrm{Ni}_{0.2}\mathrm{D}_x$ lattice ($T_0 = 77\text{--}150~\mathrm{K}$) under FMR drive, conduction $s\text{--}d$ electrons carry both equilibrium Stoner exchange polarization and a nonequilibrium spin-pumping chemical potential $\Delta\mu_s^{\mathrm{eff}}$ proportional to the FMR saturation factor $s_{\mathrm{FMR}} = h_{\mathrm{RF}}^2 / (h_{\mathrm{RF}}^2 + \Delta H^2)$ \cite{TurovPetrov1972,Overhauser1953}. Cross-relaxation via Fermi-contact hyperfine scattering transfers electron spin polarization to interstitial spin-1 deuterons according to Brillouin spin-1 statistics:
\begin{equation}
\langle M_J \rangle = \frac{2\sinh\eta_{\mathrm{DNP}}}{1 + 2\cosh\eta_{\mathrm{DNP}}}, \qquad \eta_{\mathrm{DNP}} = \frac{\hbar\gamma_D B_{\mathrm{DC}} + \xi f_{\mathrm{leak}}\Delta\mu_s^{\mathrm{eff}}}{k_B T_0}, \label{eq:dnp_c}
\end{equation}
where $\xi$ and $f_{\mathrm{leak}}$ denote the Overhauser coupling and leakage factors \cite{Overhauser1953}. Even when Suhl clamping caps $s_{\mathrm{FMR,crit}} = \Delta H / (\mu_0 M_s + \Delta H) \approx 0.053$, cryogenic ferromagnetic Overhauser pumping produces a macroscopic net alignment $|\langle M_J \rangle| > 0$ that breaks thermal spin cancellation. Nevertheless, because $\hat{\mathbf{P}}_j$ in equation~(\ref{eq:dirac_mj}) retains the $1/(M_j c^2)$ nuclear rest-mass suppression of equation~(\ref{eq:phonon_int}), this magnetoelastic Dirac $1^-$ pathway remains subordinate to the mass-independent hyperfine M1 supertransfer mechanism of equations~(\ref{eq:magnon_ham})--(\ref{eq:umag_sq}).

\ack{This work builds upon collective solid-state nuclear frameworks developed by P L Hagelstein and F Metzler \cite{Metzler2024,Hagelstein2023}. The views expressed are those of the author and do not reflect the official policy or position of DARPA or the U.S. Government.}

\funding{This work was supported by the Defense Advanced Research Projects Agency (DARPA) Defense Sciences Office (DSO) under the Project FEES and MARRS programs.}

\roles{Ryan Weed: Conceptualization, Formal analysis, Investigation, Methodology, Software, Visualization, Writing -- original draft, Writing -- review \& editing.}

\data{All analytical derivations, physical constants, and numerical scaling parameters supporting the findings of this study are included within the article and its appendices.}

\end{document}